\documentclass[twocolumn,showpacs,preprintnumbers,amsmath,amssymb,prl]{revtex4}

\usepackage{color}

\usepackage{graphicx}
\usepackage{dcolumn}
\usepackage{bm}

\usepackage[normalem]{ulem} 

\begin{document}


\title{Parallelized Quantum Monte Carlo Algorithm with Nonlocal Worm Updates}

\author{Akiko Masaki-Kato$^{1}$}
\author{Takafumi Suzuki$^{2}$}
\author{Kenji Harada$^{3}$}
\author{Synge Todo$^{1,4}$}
\author{Naoki Kawashima$^{1}$}
\affiliation{
{$^1$Institute for Solid State Physics, University of Tokyo, Chiba, Japan 277-8581}
\\
{$^2$Graduate School of Engineering, University of Hyogo, Himeji, Japan 671-2280}
\\
{$^3$Graduate School of Informatics, Kyoto University, Kyoto, Japan 615-8063}
\\
{$^4$Department of Physics, University of Tokyo, Tokyo, Japan 113-0033}
}

\begin{abstract}
Based on the worm algorithm in the path-integral representation,
we propose a general quantum Monte Carlo algorithm suitable 
for parallelizing on a distributed-memory computer by domain decomposition.
Of particular importance is its application to large lattice systems of bosons and spins.
A large number of worms are introduced and its population is controlled by a fictitious transverse field. 
For a benchmark, we study the size dependence of the Bose-condensation order parameter
of the hard-core Bose-Hubbard model with $L\times L\times \beta t = 10240\times 10240\times 16$,
using 3200 computing cores, which shows good parallelization efficiency.
\end{abstract}

\maketitle

In various numerical methods for studying quantum many body systems, 
the quantum Monte Carlo (QMC) method, in particular the worldline Monte Carlo method 
based on the Feynman path integral \cite{Suzuki}, is often used as one of the standard techniques
because of its broad applicability and exactness (apart from the controllable statistical uncertainty).
Among its successful applications, most notable are 
superfluidity in a continuous space \cite{Ceperley,Corboz},
the Haldane gap in the spin-1 antiferromagnetic Heisenberg chain \cite{Haldane,Nightingale},
and the BCS-BEC crossover \cite{VanHoucke}.
The QMC method has become more useful due to developments in both algorithms and machines.
While global update algorithms, such as loop \cite{Loop} and worm updates \cite{Prokofev},
are crucial in taming the QMC methods' inherent problem, i.e., the critical slowing-down,
the increase in computers' power following the Moore's law has been pushing up 
the attainable computation scale.
However, it is far from trivial to design the latest algorithms to benefit from the latest machines,
since the recent trend in supercomputer hardware is  ``from more clocks to more cores''; e.g.,
all top places in the TOP500 ranking based on LINPACK scores are occupied by machines 
with a huge number of processing cores \cite{Top500}.
As for the loop update algorithm, there is an efficient parallelization, 
such as the ALPS/{\footnotesize LOOPER} \cite{ALPS} code,
which now makes it possible to clarify quantum critical phenomena 
with a large characteristic length scale \cite{Harada}. 
Unfortunately, the loop update algorithm requires rather stringent conditions 
about the problems to be studied;
it is well known that the algorithm does not work for antiferromagnetic spin systems 
with external field, nor for bosonic systems with repulsive interactions.
In contrast, the worm algorithm enjoys a broader range of applicability \cite{Trotzky}.
However, the parallelization of the worm algorithm is not straightforward.
The reason is simply that the worldline configuration is updated 
by a single-point object, namely, the worm.
This fact makes the whole algorithm event-driven, hard to parallelize.
For these reasons, the parallelization of the worm algorithm has been a major challenge
from a technical point of view.

In this Letter we present a parallelized multiple-worm algorithm (PMWA) for QMC simulations.
A PMWA is generalization of the worm algorithm and it removes 
the intrinsic drawback due to the serial-operation nature 
by introducing a large number of worms.
With many worms distributed over the system, it is possible to decompose 
the whole space-time into many domains, each being assigned to a processor.
The neighboring processors send and receive updated configurations on their boundaries, 
once in every few Monte Carlo (MC) steps.
Therefore, the time required for communication can be negligible 
for sufficiently large systems.
Moreover, with PMWA we can measure an arbitrary $n$-point Green function
which is difficult in conventional worm-type algorithms when $n \ge 4$.

\begin{figure}[t]
\includegraphics[angle=0,width=8.5cm,trim= 0 0 0 0,clip]{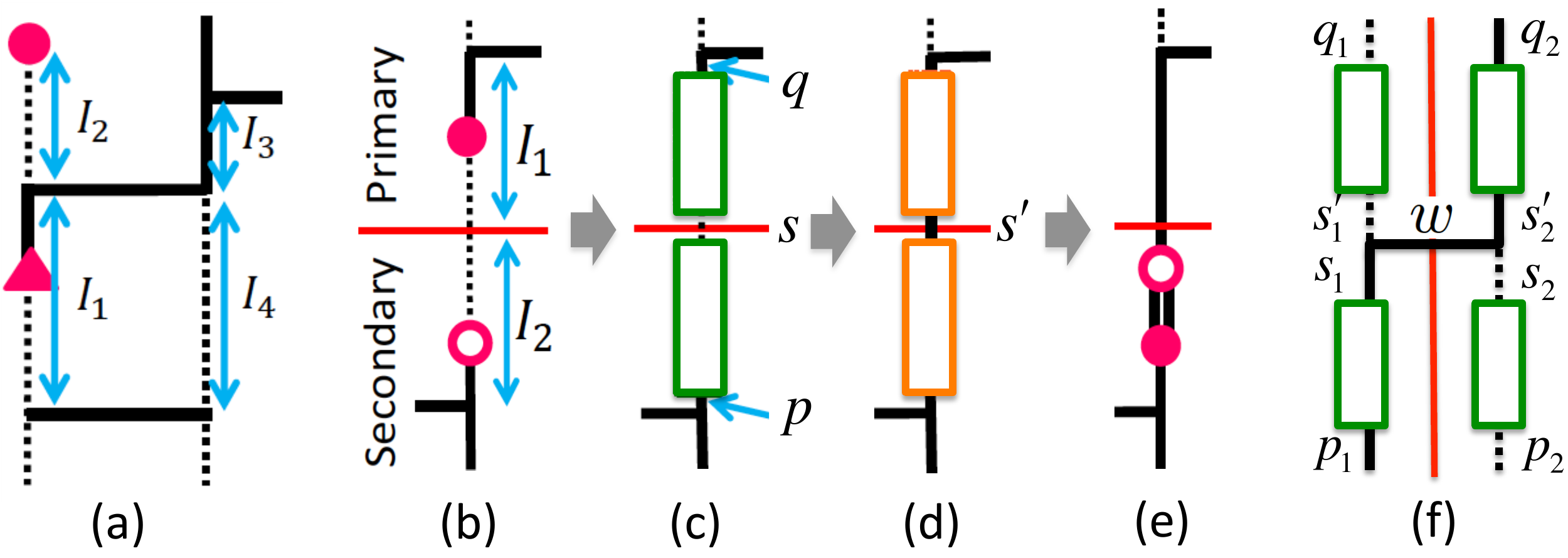}
\caption{
A vertex, its four legs, and worms (a), and creation or annihilation of worms on the temporal (b),(c),(d),(e) and on the spatial domain boundary (f).
Circles and the triangle are worms whereas horizontal lines are vertices. 
(b) The configuration before the boundary update. (The red horizontal line is the temporal domain boundary.)
(c) The initial intermediate state is $|s\rangle$.
(d) The intermediate state is updated.
(e) The final configuration compatible to the new intermediate state is generated.
(f) The vertex $w$ on the spatial domain boundary, the red vertical line.
}
\label{fig:Ptime}
\end{figure}

The algorithm described in what follows is based on the directed-loop 
implementation of the worm algorithm (DLA) \cite{DLA, Kawashima} that samples from the distribution
$W(\{\psi_k \})=\lim_{N_\tau\rightarrow\infty}\prod^{N_\tau}_{k=1}\langle \psi_{k+1}| 1-\Delta\tau {\cal H}_\eta |\psi_k\rangle$,
where $\Delta\tau=\beta/N_\tau$, $|\psi_k\rangle$ is a basis vector in some complete orthonormal basis set, 
and ${\cal H_\eta}={\cal H}-\eta Q$ is the Hamiltonian with a fictitious source term $\eta Q$ 
that generates discontinuities of worldlines, namely ``worms."
A configuration in DLA is characterized by a graph, edges and vertices, and 
state variables defined on edges in the graph.
While a vertex is represented by a point in the standard graph theoretical convention, 
in the literature of the QMC method for lattice systems it is usually represented by a short 
horizontal line connecting four vertical segments (edges) as in Fig.~\ref{fig:Ptime}.
A vertex at which the local state changes is called a ``kink.''
The update procedure of the conventional DLA consists of two phases; 
the worm phase in which the motion of the worm causes changes in the state variables,
and the vertex phase in which vertices are redistributed.
See Refs.~\cite{DLA,Kawashima} for details of these updates.
While the vertex phase in the new algorithm is just the same as the conventional DLA,
the worm phase must be modified.
In contrast to the conventional DLA, we let the worms proliferate or decrease freely 
according to the weight controlled by the parameter $\eta$.
In conventional DLA, therefore, we ``wait'' for the worms disappear to measure the observables.
(As we see below, configurations with worms are also useful in measuring off-diagonal quantities,
i.e., ``$G$-sector measurements''discussed in Refs.~\cite{Prokofev,Dorneich}.)
In the present algorithm, we estimate them instead by extrapolation to the $\eta=0$ limit.
Corresponding to this modification, the worm update is modified in two ways:
worms are created and annihilated at many places at the same time, 
and we introduce a special update procedure for the region near the boundaries.
As a result, the worm phase in the new algorithm consists of three steps: 
worm creation and annihilation, worm scattering, and a domain-boundary update.
The last step is necessary only for parallelization, 
and is not used when the program runs on a serial machine.

\textit{Worm creation and annihilation}.---Now we consider
how to assign worms on an edge or an interval $I$ separated by two vertices.
Generally, the worm-generating operator is defined as $Q_i=\sum_{i,\alpha} Q_{i,\alpha}$ 
with $Q_{i,\alpha}$ being some local operator and $i$ and $\alpha$ specifying 
the spatial position and the type of the operator, respectively.
(For example, for Bose-Hubbard model $Q_i=\sum_{\alpha=1,2}{b_{i,\alpha}}$
with the boson annihilation operator $b_{i,1}=b_i$ and the creation operator $b_{i,2}=b_i^\dagger$ 
at the site $i$.)
As is the case of the graph representation of the ${\cal H}$ term,
the probability of having $n$ worms in $I$ 
for a specified sequence of $\alpha$s, $(\alpha_1,\alpha_2,\cdots,\alpha_n)$
is given by 
$
  P_{n}^{q,p}(I,\{\alpha_k\})=((I\eta)^n/n!)\langle q|Q_{i,\alpha_n}\cdots Q_{i,\alpha_1}|p\rangle / f_{qp}(I)
$, 
where $|p\rangle$ and $|q\rangle$ are the initial and the final state of $I$ respectively, and $f_{qp}(I)\equiv\langle q| e^{I\eta Q_i} |p\rangle$.
By taking the summation over all possible sequences, 
we obtain the probability of choosing $n$ as the number of worms:
\begin{eqnarray}
P_{n}^{q,p}(I)=\frac{(I\eta)^n}{n!}\frac{\langle q|Q_i^n|p\rangle}{f_{qp}(I)}.
\label{eq:worm}
\end{eqnarray}
Once we have chosen an integer $n$ with this probability, 
we then choose a sequence of $n$ worms (or $\alpha$ s) with the probability 
$
  \langle q|Q_{i,\alpha_n}\cdots Q_{i,\alpha_1}|p\rangle / \langle q|Q_i^n|p\rangle
$.
After having chosen $n$ and the sequence the $n$ operators in this way, 
we choose $n$ imaginary times uniform randomly in $I$ and
place the $n$ worms there according to the sequence selected above.

While the present algorithm is quite general, 
let us consider the hard-core Bose-Hubbard model for making the discussion concrete,
for which the algorithm becomes simple.
In this case, Eq.~(\ref{eq:worm}) is nonvanishing only if
$n$ is even for $|q\rangle=|p\rangle$ or $n$ is odd for $|q\rangle\neq |p\rangle$,
and in either case the sequence that has nonvanishing weight is unique,
alternating between $b$ and $b^{\dagger}$.
We here introduce a variable $\sigma$, which specifies the ``parity" of 
the number of worms in an interval $I$; 
$\sigma=0$ when $|q\rangle=|p\rangle$ and $\sigma=1$ when $|q\rangle\neq|p\rangle$ .
For each parity, the probability, Eq.~(\ref{eq:worm}), becomes the following simple form
analogous to the Poisson distribution,
$
P^\sigma_n(I)=((I\eta)^n/n!)(1/f_\sigma(I)), \quad \{ n\in \mathbb{N} | n \ \text{mod}\  2 = \sigma \}.
$
Here $f_{\sigma}(I)=\cosh(I\eta)$ for $\sigma=0$ and  $\sinh(I\eta)$ for $\sigma=1$.

\textit{Worm scattering}.---Now we consider how we let the worms move around.
Note that every worm has the direction, up or down,
and according to the nearest object in this direction,
different action should be taken.
If it is another worm, then we simply change the direction of 
the worm and do not change its location.
If it is a vertex, we let the worm scatter there.
Below we discuss how this scattering procedure should be done.

Suppose that a worm is on the $i$th leg of the vertex.
Here a leg is an interval delimited by the vertex in question on one end, 
and by another vertex or another worm on the other.
Then, with probability $P_{{\rm enter}} \equiv L_{{\rm min}}/L_i$, we let it enter the vertex,
where $L_{{\rm min}}$ is the length of the shortest of the four legs connected to the vertex
[Fig.~\ref{fig:Ptime}(a)].
Otherwise, we let it turn around without changing its position.
If it enters the vertex, it chooses the out-going leg $j$ with probability
$P_{{\rm scatter}} = w_{ji}/w_i$, where $w_i$ is the weight of the state with the worm on the $i$th leg. 
This is the usual scattering probability in DLA.  
Here, $w_{ji}$ satisfies two equations, $w_{ij}=w_{ji}$ and $w_i=\sum_lw_{li}$,
where $l$ runs over all leg indices.
Finally, the imaginary time of the worm is chosen uniform randomly in the interval $L_j$.
These procedures define the following transition probability:
\begin{equation}
\label{eq:DB}
p_{i\rightarrow j} = \frac{L_{{\rm min}}}{L_i} \frac{w_{ji}}{w_i}\frac{\Delta\tau}{L_j}.
\end{equation}
It is obvious that this transition probability satisfies 
the detailed-balance condition (to be more precise, the time-reversal symmetry condition in the present case),
$p_{i\rightarrow j} w_i = p_{j\rightarrow i} w_j$.
The number of worm scatterings in a MC step is chosen so that 
every part of the space-time is updated roughly once on average.

\textit{Boundary-configuration update}.---In the parallelized calculation,
we decompose the whole space-time into multiple domains.
Then, there are two special cases in the worm scattering discussed above;
the case where the worm tries to enter a vertex connecting two domains
(spatial domain boundary),
and the case where the worm tries to go out of the current domain and 
enters another (temporal domain boundary).
In these two cases, the worm
is bounced by the vertex or the boundary with probability one. 
This treatment obviously satisfies the detailed-balance condition, but
it breaks the ergodicity.
In order to recover the ergodicity, 
we carry out the special update procedure described below 
in the region near the boundary at every MC step.

Figures \ref{fig:Ptime}(b)-(e) show the update procedure of a temporal boundary 
that has two ``legs," $I_1$ and $I_2$ [Fig.~\ref{fig:Ptime}(b)], 
ending with states $|q\rangle$ and $|p\rangle$, respectively. 
We choose the processor taking care of the upper domain as the ``primary" 
and let it execute operations for updating the pair.
The current local state just at the boundary is $|s\rangle$ [Fig.~\ref{fig:Ptime}(c)].
Then, the new state $|s'\rangle$ is chosen with the probability,
\begin{eqnarray}
\label{eq:Para}
P_{\rm dom}^{s'}=\frac{f_{qs'}(I_1)f_{s'p}(I_2)}{f_{qp}(I)}
\end{eqnarray}
[Fig.~\ref{fig:Ptime}(d)].
Once $|s'\rangle$ has been chosen, we can regenerate all worms in $I_1$ and $I_2$ 
with Eq.~(\ref{eq:worm}) as discussed previously [see Fig.~\ref{fig:Ptime}(e)].
For hard-core bosons, for example, Eq.~(\ref{eq:Para}) is explicitly rewritten as
$
P_{\rm dom}^{\sigma'_1,\sigma'_2}$ $=(1+\tanh^{S(\sigma'_1)}(I_1\eta)\tanh^{S(\sigma'_2)}(I_2\eta))^{-1}
$,
where $\sigma_1$ and $\sigma_2$ are the parities of $I_1$ and $I_2$, respectively, and $S(0)=1$ and $S(1)=-1$. 

The update procedure of the interdomain vertex is similar 
to that of the temporal boundary discussed above,
although there are four intervals involved in this case instead of two.
Suppose we have a vertex with four legs bounded with the ending states $|p_1\rangle$ and $|p_2\rangle$ 
below the vertex, and $|q_1\rangle$ and $|q_2\rangle$ above the vertex.
Now, the new state variables $s_1, s_2, s'_1, s'_2$ at the roots of the four legs as shown in Fig.~\ref{fig:Ptime}(f) 
are stochastically selected according to the product of the vertex weight $w$ and the leg weight $f$,
\begin{eqnarray}
W^{p_1p_2s_1s_2}_{s'_1s'_2q_1q_2}=f_{q_1s'_1}(I_3)f_{q_2s'_2}(I_4)w^{s_1s_2}_{s'_1s'_2}f_{s_1p_1}(I_1)f_{s_2p_2}(I_2),
\label{eq:ParaSpace}
\end{eqnarray}
where $w^{s_1s_2}_{s'_1s'_2}$ is $\langle s'_1,s'_2 | H_{\rm pair} | s_1, s_2 \rangle$,
previously referred to as $w_i$ in Eq.~(\ref{eq:DB}) with
$i$ representing the set root states $s_1,s_2,s'_1,s'_2$.
Once the new root states have been selected, the rest of the task is the
same as the temporal boundary update; i.e., we regenerate worms on the four legs.
These tasks are carried out by the primary processor that
takes care of the ``left"-hand side of the vertex.

\textit{Pseudo code}.---We summarize all of the procedure described above in the form of a pseudo code.
The task of a processor $\nu$ in a MC step is as follows,

\noindent(Step 1) Send to and receive from the neighboring processor
the ending states of the intervals on the temporal boundary.
For each one of the intervals of which $\nu$ is primary, 
select the intermediate state stochastically with the probability Eq.~(\ref{eq:Para}).
Send and receive the updated intermediate states.

\noindent(Step 2) Send to and receive from the neighboring processor
the ending states of the legs of the vertices on the spatial boundary.
For each one of the vertex of which $\nu$ is primary,
select the states at the roots of the legs stochastically 
with the weight Eq.~(\ref{eq:ParaSpace}).
Send and receive the updated root states.

\noindent(Step 3) As in the conventional DLA, erase all vertices without a kink on it, 
and place new vertices with the density proportional to the corresponding diagonal 
matrix element of the Hamiltonian.

\noindent(Step 4) For each interval $I$ delimited by the vertices or the domain boundaries, 
erase all the worms, generate an integer $n$ with the probability Eq.~(\ref{eq:worm}),
generate a sequence of $n$ operators, and place them uniform randomly on $I$.
Also choose the direction of each worm with probability 1/2.

\noindent(Step 5) 
For every worm, if the nearest object ahead is a vertex
that is not on a boundary, let it enter the vertex with the probability $P_{{\rm enter}}$,
let the worm scatter there with the probability $P_{{\rm scatter}}$,
and choose the imaginary time uniform randomly on the final leg.
Otherwise, reverse its direction without changing its position.

\noindent(Step 6) Repeat Step 5 $N_{{\rm cycle}}-1$ more times, and perform measurements. 
This concludes the MC step.

\textit{Benchmark}.---We apply the algorithm to the hard-core Bose-Hubbard model in the square lattice.
The model we consider here is defined as
\begin{eqnarray}
{\cal H}=-t\sum_{\langle i,j\rangle}b^\dagger_ib_j+V\sum_in_{i}n_{j}-\mu\sum_i(n_i+n_j),
\label{eq:BH}
\end{eqnarray}
where $\mu$ is the chemical potential and $V$ denotes the nearest-neighbor interaction.
In the PMWA, we simulate the Hamiltonian ${\cal H}_\eta$ to generate multiple worms, 
then we extrapolate the QMC results to the $\eta=0$ limit.
The extrapolation rule is given by the expansion of the physical quantity 
in a power series of $\eta$ where $\eta$ is small.
For example, the coefficient of the first order term of the energy is as follows:
\begin{equation}
\label{eq:Eng}
\left. \frac{\partial \langle {\cal H} \rangle_\eta}{\partial\eta}\right|_{\eta\rightarrow 0} 
= - \langle Q\rangle_0 + \beta\langle {\cal H} Q\rangle_0 - \beta\langle{\cal H}\rangle_0\langle Q\rangle_0,
\end{equation}
where $\langle \cdots \rangle_0$ is the mean value with respect to 
the nonperturbed Hamiltonian (\ref{eq:BH}).
When we choose $Q$ to be a measure of the spontaneous symmetry breaking,
as we do below, the right-hand side of Eq.~(\ref{eq:Eng}) is always 0 for a finite system,
making the ${\cal O}(\eta)$ term in $\langle {\cal H}\rangle$ vanish.
\begin{figure}[htpb]
\includegraphics[angle=0,width=8.4cm,trim= 0 0 0 0,clip]{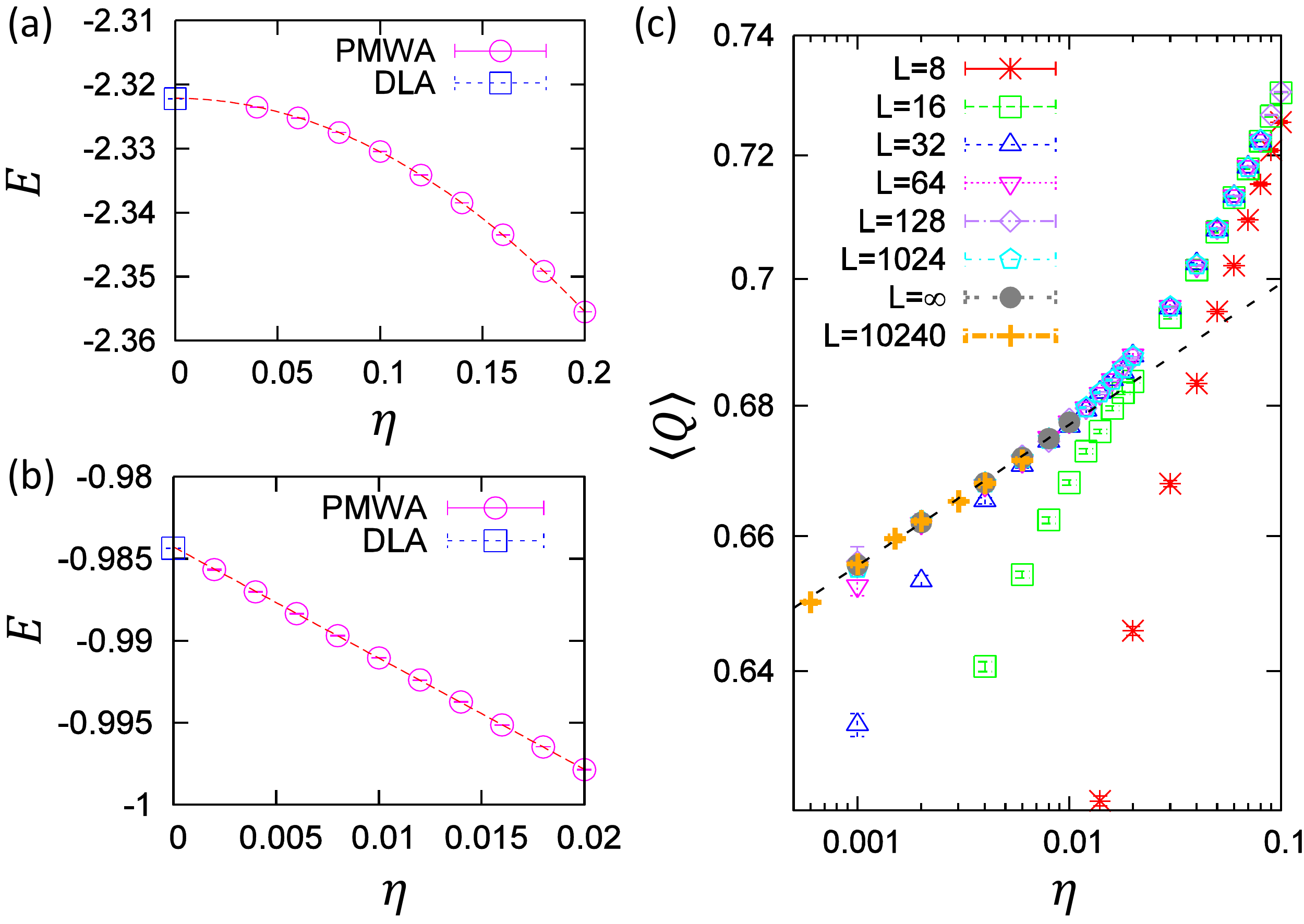}
\caption{
Energy $E$ and order-parameter $Q$ as a function of the source field $\eta$.
(a) and (b): Energy at fixed system size, temperature, and repulsive interaction ($L=128, \beta t = 16, V/t=3.0$).
The chemical potential is $\mu / t = 4.2$ (CS phase) for (a), and $\mu/t= 1.2$ (SF phase) for (b).
The dashed curves represent a quadratic fitting for (a) and a linear fitting for (b).
(c): Double logarithmic plot of $\langle Q\rangle$ at $\beta t = 16$, $\mu / t = 1.2$ and $V/t = 3.0$.
The dashed line is the power-law fitting with $L=\infty$ data.
}
\label{fig:Eng}
\end{figure}
This result leads us to quadratic extrapolation $-2.32216(4)$,
which shows good agreement with the conventional DLA result $-2.32222(2)$ in the checkerboard solid (CS) phase [Fig.~\ref{fig:Eng}(a)].
In contrast, in the superfluid (SF) phase $\langle Q\rangle$ is finite
in the thermodynamic limit at zero temperature.
Even for finite systems at finite temperature, 
the deviation from the thermodynamic behavior at $T=0$ appears only in very 
small $\eta$ and practically not observed in large systems for which 
parallelization is necessary.
It allows us to extrapolate the energy linearly at low temperatures as we see in Fig.~\ref{fig:Eng}(b)
in which values are $-0.98431(1)$ by the PMWA and $-0.98434(2)$ by DLA.
By closer inspection, however, we can also estimate the continuously varying scaling exponent,
characteristic of the two-dimensional systems at finite temperature.
Below we demonstrate that the present method can produce the off-diagonal order parameter, 
namely, the Bose-Einstein condensate (BEC) order parameter $\langle Q\rangle$.
The procedure for measuring this quantity and an arbitrary multipoint Green function,
i.e. $G$-sector measurements, is shown in the Supplemental Material \cite{Supplement}.
Figure \ref{fig:Eng}(c) shows the numerical results for systems ranging 
from $L=8$ to $10240$ at fixed $\beta t=16$, 
which is much larger than a single processor can accommodate.
We also present the result of extrapolation to the infinite $L$ limit for each value of $\eta$ based on the results of $L\leq 1024$.
The $L=10240$ results calculated by using 3200 CPU cores agree well with the extrapolation.
The dashed line is the power law fitting from which we can read the magnetization critical exponent $1/\delta$.
\begin{figure}[htpb]
\includegraphics[angle=0,width=8.4cm,trim= 0 0 0 0,clip]{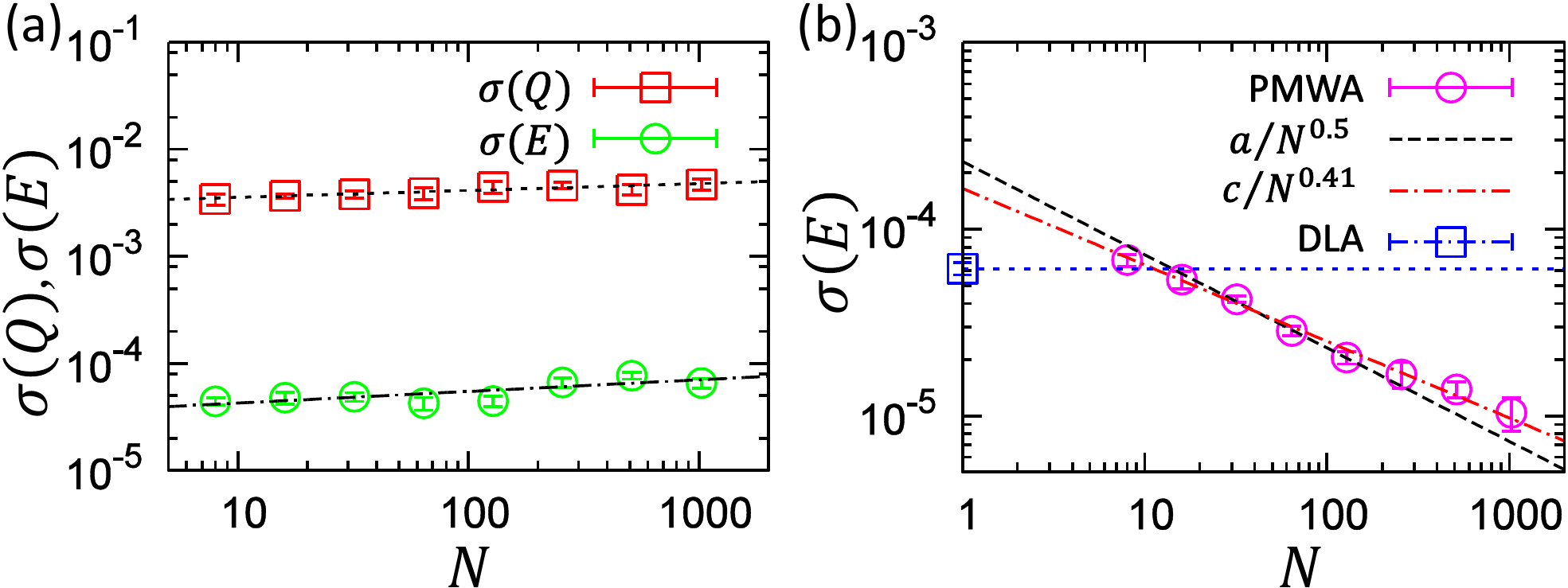}
\caption{
The estimated standard deviation (error) as functions of the number of domains $N$ in a SF state ($\mu/t = 1.2, V/t = 3.0$).
(a) With fixed number of MC sweeps. ($L=\beta t=128$, $\eta=0.002$),
(b) With fixed wall-clock time of $10^4$ seconds (not including thermalization).
($L=256$ and $\beta t = 16$, $\eta=0.004$).}
\label{fig:Q}
\end{figure}

It is in general possible that the domain boundaries hinder the
propagation of the locally equilibrated region, and cause a slowing-down.
In order to see the seriousness of this effect,
we estimated the standard error,
i.e., 1 standard deviation of the expected distribution,
of the mean values of the order parameter $\sigma(Q)$
and that of the energy $\sigma(E)$ in SF phase
as a function of the number of domains $N$ in Fig.~\ref{fig:Q}(a).
We measured the quantities at every MC sweep, and the averages are taken over the same number of MC sweeps.
Though the number of worms decrease with increasing $N$,
in all of our presented simulation the average number of worms in each domain is from $O(10)$ to $O(100)$ for used $\eta$, 
and the probability of finding an empty domain is very low.
We find a weak $N$ dependence of $\sigma$, which is empirically described as $\sigma \propto N^{0.09}$.
We here emphasize that the PMWA is also efficient from the 
technical point of view; i.e., each processor has to communicate
only with its neighbors, and the amount of transmitted information
is proportional to the area of only the interface.
This property is manifested in the good parallelization performance of our algorithm and code.
For the so-called ``weak scaling" performance, with the system size being proportional to the number of processors,
see the Supplemental Material \cite{Supplement}.
Figure~3(b) suggests good ``strong scaling" performance, with the fixed system size and increasing number of processors.
Specifically, it shows the standard error as a function of $N$ with both the system size and the  elapsed time (``wall-clock" time) fixed.
It plainly shows that we can achieve higher accuracy by employing more processors.
The $N$ dependence of the error is described (again empirically) by $N^{-0.41}$ which is slightly worse than the ideal dependence $N^{-0.5}$.
The small difference in the exponent $0.09$ comes from the slight increase observed in Fig.~3(a).

We have presented a PMWA, a new parallelizable QMC algorithm, 
which can treat extremely large systems. 
We have applied it to hard-core bosons and observed high parallelization efficiency.
The multibody correlation function should be computed relatively easily
in the new algorithm.
In addition, the PMWA can be extended in several ways.
For example, ``on-the-fly'' vertex generation \cite{Kato,Pollet}, in which vertices are 
generated only in the immediate vicinity of the worms, is possible.
Another extension may be the ``wormless'' algorithm.
Obviously, the boundary update in terms of the parity of the number of worms rather 
than worms themselves can be used also for updating bulk regions.
By doing so, we can altogether get rid of worms.
These extensions will be discussed elsewhere \cite{FuturePaper}.
The source code of our program will be released in Ref.~\cite{DSQSS} in the near future.

We would like to thank H. Matsuo, H. Watanabe, T. Okubo, R. Igarashi, and T. Sakashita for many helpful comments.
This work was supported by CMSI/SPIRE, the HPCI System Research project (hp130007),
and Grants-in-Aid for Scientific Research No. 25287097.
Computations were performed on computers at the Information Technology Center of the University of the Tokyo, 
and at Supercomputer Center, ISSP.

\subsection{Supplemental Material}
\subsection{S-I. Procedure of measuring the BEC order parameter and the multipoint Green function}
We show here how to measure the BEC order parameter in our QMC method.
The expectation value of an operator $Q_\alpha(X=\tau,{\bm r_i})\equiv{\rm e}^{\tau H}Q_{i,\alpha}{\rm e}^{-\tau H}$ is expressed as follows,
\begin{eqnarray}
\langle Q_\alpha(X)\rangle_\eta &=& \frac{{\rm Tr}\:T_{\tau}\left[Q_\alpha(X){\rm e}^{-\beta(H-\eta Q)]}\right]}{{\rm Tr}\:{\rm e}^{-\beta(H-\eta Q)}}\nonumber,
\end{eqnarray}
where $T_\tau$ is the time-ordering operator.
The numerator is 
\begin{align}
{\rm Tr}\:T_{\tau}&\left[Q_\alpha(X){\rm e}^{-\beta(H-\eta Q)]}\right]\nonumber\\
=&{\rm Tr}\left({\rm e}^{-\beta H}\sum_n\frac{\eta^n}{n!}\int^\beta_0d\tau_1\cdots d\tau_n \right.\nonumber\\
&\Biggl. \qquad \quad T_\tau \left[Q_\alpha(X)Q_{\alpha_n}(X_n)\cdots Q_{\alpha_1}(X_1)\right]\Biggr)\nonumber\\
=&\frac{1}{\eta}{\rm Tr}\left({\rm e}^{-\beta H}\sum_n\frac{\eta^{n}}{n!}\int^\beta_0d\tau_1\cdots d\tau_{n}\right. \nonumber\\
&\left. \qquad \quad T_\tau \left[Q_{\alpha_{n}}(X_{n})\cdots Q_{\alpha_1}(X_1)\right]\sum^{n}_{k=1}\delta(X_k=X)\right).\nonumber
\end{align}
Thus, we obtain
\begin{eqnarray}
\langle Q_\alpha(X)\rangle_\eta &=& \frac{1}{\eta}\langle \rho_\alpha(X)\rangle_{\eta,{\rm MC}}.\nonumber
\label{eq:Q}
\end{eqnarray}
The symbol $\rho_\alpha(X)$ denotes the MC observable 
for the density of worms at $X$.
It is defined as
\begin{eqnarray}
\rho_{\alpha}(X) & = & \frac{\eta\langle q|e^{I_2\eta Q}Q_\alpha e^{I_1\eta Q}|p \rangle}{\langle q|e^{I\eta Q}|p \rangle}\nonumber\\
&=& \frac{\eta\sum_{s,s'} f_{qs'}(I_2) \langle s' | Q_{\alpha} | s \rangle f_{sp}(I_1)}{f_{qp}(I)}
\quad (X\in I),\nonumber
\label{eq:PII}
\end{eqnarray}
where $I$, the interval on which X is located, is split into $I_1$ and $I_2$ at $X$.
The final and initial states of $I$ are $q$ and $p$ respectively.
Using $\rho_{\alpha}(X)$, the MC observable of the arbitrary $n$-body Green function is simply expressed as follows:
\begin{eqnarray}
\left\langle \prod_{k=1}^n Q_{\alpha_k}(X_k)\right\rangle_\eta=\frac{1}{\eta^n}\left\langle \prod_{k=1}^n \rho_{\alpha_k}(X_k)\right\rangle_{\eta,{\rm MC}},\nonumber
\label{eq:G}
\end{eqnarray}
with the exceptions of the cases where multiple $X_k$ fall on the same
interval.
Details such as this will be discussed in our upcoming paper.
\subsection{S-II. Weak-scaling acceleration efficiency}
\begin{figure}[b]
\includegraphics[angle=0,width=6cm,trim= 0 0 0 0,clip]{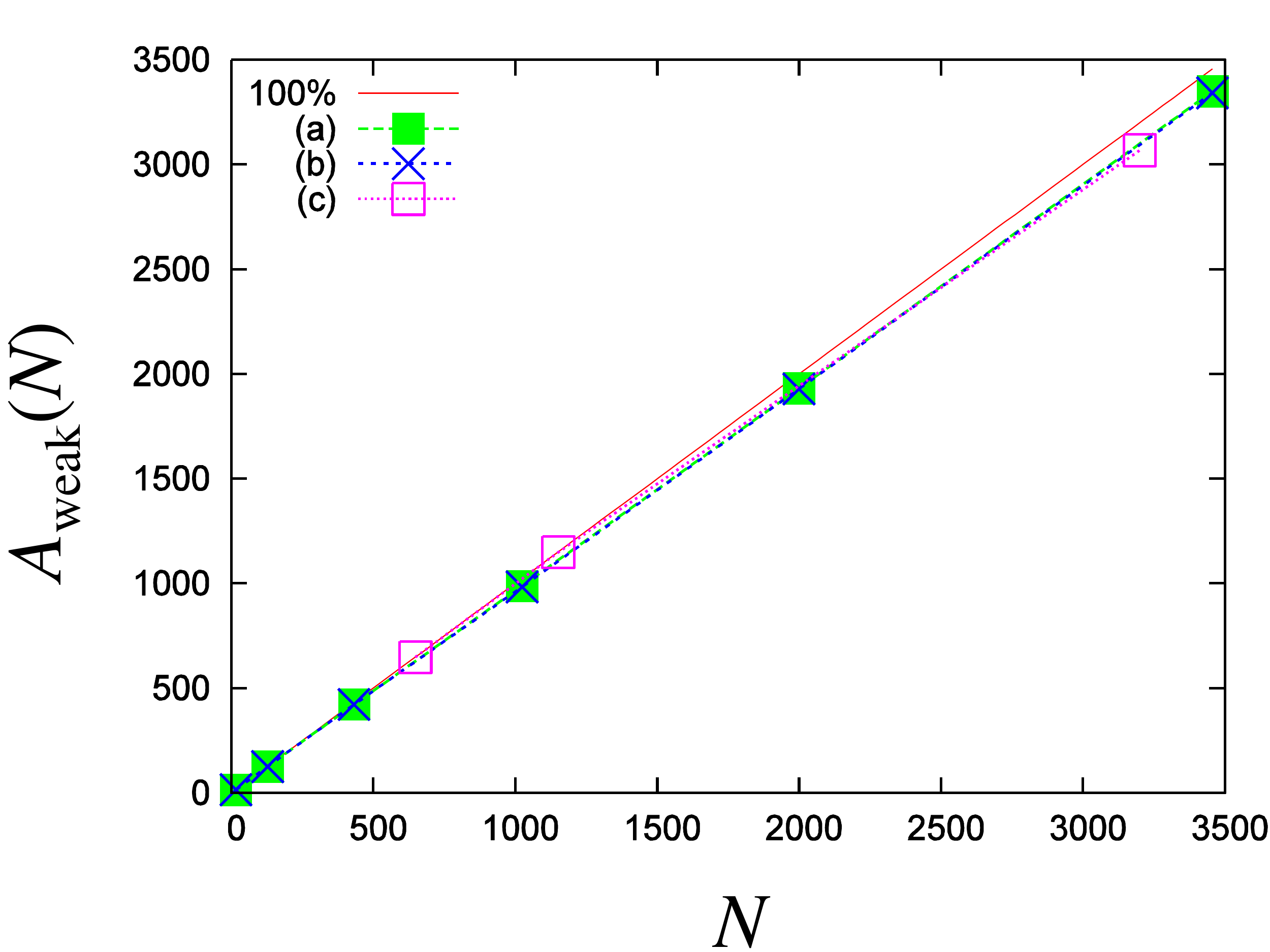}
\caption{\label{fig:PE}
Weak-scaling acceleration efficiency of hard-core Bose-Hubbard models in the square lattice.
(a) Superfluid phase with $\mu=-0.2t$, $V=3.0t$, $\eta=0.06$. Here, $\Delta L=16$, $\Delta\beta t=8$.
(b) Checkerboard solid phase with $\mu=5.2t$, $V=3.0t$, $\eta=0.06$. Here, $\Delta L=16$, $\Delta\beta t=8$.
(c) Superfluid phase with $\mu=1.2t$, $V=3.0t$, $\eta=0.04$. Here, $\Delta L=256$, $\Delta\beta t=8$.
}
\end{figure}
The ``weak scaling'' acceleration efficiency is defined as $A_{\rm weak}(N) \equiv NT_1/T_{N}$, where $T_{N}$ is the elapsed
computational time by using $N$ processors for a system with $N$
domains which have the fixed domain size as $\Delta V=\Delta\beta\Delta L^d$.
When $N \equiv N_{\beta}\times {N_L}^d$, the total size of a system with $N$ processors is $\beta \times L^d$,
where $\beta\equiv N_{\beta} \Delta\beta, L \equiv N_L \Delta L$.
Figure~\ref{fig:PE} shows results of $A_{\rm weak}(N)$ for hard-core Bose-Hubbard models (defined in our main text) in the square lattice.
We tried various decomposition pairs with $N=N_\beta\times N_L\times N_L$ 
where $(N_\beta, N_L)=(4, 2)$, $(8, 4)$, $(12, 6)$, $(16, 8)$, $(20, 10)$, $(24,12)$ in Fig.~\ref{fig:PE}(a) and (b),
and $(2, 18)$, $(2, 24)$, $(2, 40)$ in  Fig.~\ref{fig:PE}(c).
In our calculation on FUJITSU PRIMEHPC FX10,
we found that for large $N$ the efficiency is slightly less than 1.
This slowing down may be caused
by the ``load balance'' problem or by the information passing between processors.
In all of our presented simulation, 
we confirmed that the amount of CPU time consumed by the information passing is
negligibly small even when the number of processors is large
(according to FUJITSU's profiler data, $\sim 2.6\%$ of the total computational cost for the information passing and
$\sim 13.6\%$ of the total computational cost for idle time when $N=1024$).
Therefore, the main source of the deviation from the ideal curve is 
the load-balancing, i.e., 
the amounts of computational tasks for processors become uneven 
causing some processors to finish their tasks earlier than the others and be idle.
However, the efficiency only decreases by $\sim 3\%$ even when $N\gtrsim 1024$.
The efficiency in a superfluid phase and a checkerboard solid phase turned out almost the same.


\end{document}